\documentclass[11pt]{article}
\usepackage[margin=1.05in]{geometry}
\usepackage{amsmath,amssymb,amsthm}
\usepackage[T1]{fontenc}
\usepackage{graphicx}
\usepackage{booktabs}
\usepackage{xcolor}
\usepackage[colorlinks=true,allcolors=blue]{hyperref}
\hypersetup{
  pdftitle={Mass Dependence of Araki Relative Entropy through Modular Theory},
  pdfauthor={M. S. Guimaraes, I. Roditi, S. P. Sorella},
  pdfsubject={Revised version V39, 8 September 2026},
  pdfkeywords={Araki relative entropy, modular theory, standard subspaces,
    Bisognano-Wichmann, coherent states, Tomita-Takesaki, wedge localization}
}

\theoremstyle{remark}
\newtheorem{remark}{Remark}

\newcommand{\Hh}{\mathcal{H}}
\newcommand{\im}{\operatorname{Im}}
\newcommand{\re}{\operatorname{Re}}
\newcommand{\dom}{\operatorname{Dom}}
\newcommand{\ran}{\operatorname{Ran}}
\newcommand{\W}{\mathrm{W}}
\newcommand{\sinc}{\operatorname{sinc}}

\title{\textbf{Mass Dependence of Araki Relative Entropy through Modular Theory}}
\author{M.~S.~Guimaraes$^{1}$,\quad I.~Roditi$^{2}$,\quad S.~P.~Sorella$^{1}$\\[2.5mm]
{\footnotesize $^{1}$UERJ -- Universidade do Estado do Rio de Janeiro, Instituto de F\'isica, Departamento de F\'isica Te\'orica,}\\[-0.4mm]
{\footnotesize Rua S\~ao Francisco Xavier 524, 20550-013, Maracan\~a, Rio de Janeiro, Brazil}\\[-0.4mm]
{\footnotesize \texttt{msguimaraes@uerj.br}\ \ \texttt{silvio.sorella@fis.uerj.br}}\\[1.5mm]
{\footnotesize $^{2}$CBPF -- Centro Brasileiro de Pesquisas F\'isicas, Rua Dr.~Xavier Sigaud 150, 22290-180, Rio de Janeiro, Brazil}\\[-0.4mm]
{\footnotesize \texttt{roditi@cbpf.br}}}
\date{}

\begin{document}
\maketitle

\begin{abstract}
\noindent Building on the established one-particle formula for the Araki relative entropy of coherent states \cite{CLR,BCD,Longo}, we study how its value acquires a nontrivial dependence on the mass of the scalar field. For a localized vector $h$ belonging to the standard subspace $H_m$ of the one-particle Hilbert space, the known quadratic-form expression is: $S_{H_m}(h)=-\langle h,\log\delta_{H_m}\,h\rangle$.
Our contribution is to construct explicitly a mass-indexed family of vectors of the wedge standard subspace on which this expression is evaluated. The mass-shell map $h_m=E_mf$ organizes four structural conditions on rapidity representatives---on-shell dependence, a controlled massless boundary value, decay for large real rapidity, and Bisognano--Wichmann strip analyticity---and we exhibit an entire rapidity wave function, built from a doubled light-cone phase, two sinc factors, and a Gaussian pair, that satisfies them together with the sharp localization criterion: Hardy-type $L^2$ control throughout the Bisognano--Wichmann strip and the exact Tomita boundary relation. The family therefore belongs to $H_m(\W_R)$ for every $m>0$, and its Araki relative entropy is finite and strictly positive, with an exact spectral representation that makes positivity manifest. The entropy is strongly suppressed at large mass, attains a maximum at intermediate mass in $1+1$ dimension, and converges to a finite value along the modular flow as $m\to0^+$. The construction extends fiberwise to $1{+}d$ dimensions through the transverse mass.
\end{abstract}

\section{Introduction and relation to previous work}\label{sec:intro}

Relative entropy in algebraic quantum field theory is simultaneously an information-theoretic quantity and a modular-theoretic invariant. Araki's definition places it naturally in the standard form of a von Neumann algebra, while the Bisognano--Wichmann theorem identifies the modular flow of a wedge algebra with Lorentz boosts. For coherent excitations of free fields, these two structures meet particularly cleanly: the many-body relative entropy can be reduced to a quadratic expression at the one-particle level.

Let us begin by recalling the historical development. Ciolli, Longo and Ruzzi formulated the entropy of vectors relative to standard real subspaces and related it to coherent excitations of free fields \cite{CLR}; Casini, Grillo and Pontello derived coherent-state relative entropy directly from Araki's formula in free field theory \cite{CGP}; and Bostelmann, Cadamuro and Del Vecchio subsequently gave a general CCR treatment in terms of single-particle modular data \cite{BCD}. The modular-localization framework in which standard subspaces are constructed from positive-energy Poincar\'e representations goes back, in the form used here, to Brunetti, Guido and Longo \cite{BGL}. We therefore take the one-particle entropy formula as established input. The question pursued here is the structural question of how a scalar mass can re-enter after the isolated wedge modular flow has become universal, i.e.\ the same for every mass $m>0$. The unitary identification of $\Hh_m$ with $L^2(\mathbb R,d\theta)$ makes its generator the fixed operator $k_\W=2\pi D$, independent of $m$, where $D=-i\partial_\theta$. This statement concerns only the isolated modular pair.

This perspective also clarifies the relation with our earlier work in which the mass dependence was studied by choosing smooth spacetime test functions and evaluating smeared Pauli--Jordan distributions: numerically for vacuum--coherent pairs in \cite{GRSV}, for pairs of coherent states in \cite{CGRS}, and across $1{+}1$, $1{+}2$, and $1{+}3$ dimensions in \cite{Predecessor}. That representation is physically transparent, but in higher spacetime dimension the relevant oscillatory integrals can become numerically delicate. The present formulation is therefore best viewed as a continuation and conceptual reorganization of that program: instead of beginning with a spacetime smearing and then extracting modular information, we begin with the standard subspace and its Tomita--Takesaki data and ask where mass can enter at all. Relative to those direct predecessors, the new element offered here is an explicit mass-indexed family that lies in the wedge standard subspace, so that its finite, strictly positive entropy curve is an actual Araki relative entropy.

There is an important conceptual translation between the two descriptions. In the spacetime-test-function formulation the origin of the mass dependence is manifest: the smeared Pauli--Jordan distribution depends explicitly on the mass through the free-field dispersion relation and its mass-dependent kernel. Passing to the Bisognano--Wichmann modular description can make this explicit parameter seem to disappear, because the modular flow of one isolated Rindler wedge is geometrically the boost flow and is mass-independent after rapidity identification. This does not mean that a physical parameter of the theory has been lost. Rather, the mass dependence has changed its location in the description. It is encoded in the mass-indexed localization data---in the family $m\mapsto H_m(\W)$, and, more intrinsically, in the translated net $a\mapsto H_m(\W+a)$ and in the corresponding mass-dependent coherent vectors. Thus the spacetime and modular formulations display the same physical dependence in different coordinates.

A technical ingredient of the one-particle formula is the cutting projector $P_H$. Because its modular expression involves an unbounded real-linear operator and the inverse of $1-\delta_H$, Appendix~\ref{app:cutting} gives a deliberately elementary algebraic derivation on an explicitly stated spectral core. The appendix is pedagogical: it exhibits the geometric action of the formula; the closed-operator statement itself is quoted from the standard-subspace literature \cite{CLR}.

The mass-shell restriction map $E_m$ provides the constructive bridge back to spacetime. For a fixed test function, the mass dependence carried explicitly by the Pauli--Jordan distribution is transferred to the one-particle vector $h_m=E_mf$. Thus a mass-independent isolated-wedge modular generator is fully compatible with a nontrivial mass-dependent Araki relative entropy. The mass dependence studied below is carried by a family of vectors organized by four structural conditions induced by this mass-shell construction.

The present work turns this observation into a constructive result. The four conditions---on-shell variables, a controlled massless boundary value, decay for large real rapidity, and Bisognano--Wichmann strip analyticity---organize the search; the sharp membership criterion for the wedge standard subspace is the classical strip characterization given below (Hardy-type $L^2$ control in the strip together with the Tomita boundary relation). Our main result is an explicit entire family that satisfies all of it: a doubled light-cone phase whose interior-strip decay exactly cancels the growth of two sinc factors, and a Gaussian pair that supplies real-axis decay and transforms correctly under the strip shift $\theta\mapsto\theta+i\pi$. We then establish localization, derive an exact spectral proof of positivity for the resulting entropy, and compute the mass curve. Transverse momentum gives the organized $1{+}d$ extension.

The paper is organized as follows. We first review standard real subspaces, coherent states, Araki relative entropy, the cutting projection, and the one-particle entropy formula, and we record the rapidity characterization of the wedge subspace. We then show explicitly how a fixed spacetime preparation produces the mass-indexed modular vector $h_m=E_mf$. The subsequent sections present the explicit localized family in $1{+}1$ dimensions---its localization, the positivity and spectral representation of its entropy, and the computed mass curve---and the transverse-mass extension to $1{+}d$. The appendices collect the cutting-projector derivation, the elementary second-quantized derivation of the one-particle formula, and the operational meaning of relative entropy.

\section{Standard real subspaces and Tomita--Takesaki data}\label{sec:standard}

Let $\Hh$ be a complex Hilbert space. A closed real-linear subspace $H\subset\Hh$ is \emph{standard} if
\begin{equation}
\overline{H+iH}=\Hh,\qquad H\cap iH=\{0\}. \label{eq:standard}
\end{equation}
On the dense domain $H+iH$ define the antilinear Tomita--Takesaki operator
\begin{equation}
s_H(h_1+ih_2)=h_1-ih_2,\qquad h_1,h_2\in H. \label{eq:tomita}
\end{equation}
It is closed, satisfies $s_H^2=1$ on its domain, and has polar decomposition
\begin{equation}
s_H=j_H\delta_H^{1/2}, \label{eq:polar}
\end{equation}
where $j_H$ is antiunitary and $\delta_H$ is positive self-adjoint. The standard subspace is recovered as
\begin{equation}
H=\{h\in\dom s_H:\ s_Hh=h\}. \label{eq:recover}
\end{equation}
The symplectic complement is
\begin{equation}
H'=\{k\in\Hh:\ \im\langle k,h\rangle=0\ \text{for all}\ h\in H\}, \label{eq:sympcompl}
\end{equation}
and modular theory gives
\begin{equation}
j_HH=H',\qquad \delta_H^{it}H=H. \label{eq:modularaction}
\end{equation}
We shall assume factoriality whenever the cutting decomposition is used:
\begin{equation}
H\cap H'=\{0\}. \label{eq:factorial}
\end{equation}
In the CCR construction, this is the one-particle counterpart of the factor condition for the associated von Neumann algebra; it ensures uniqueness of the real $H+H'$ decomposition on its natural domain.

\subsection{Weyl algebra and coherent states}\label{sec:weyl}

The bosonic Weyl algebra over $\Hh$ is generated by unitaries $W(f)$ satisfying
\begin{equation}
W(f)W(g)=e^{-i\im\langle f,g\rangle}W(f+g),\qquad W(f)^*=W(-f). \label{eq:weyl}
\end{equation}
By means of the bicommutant theorem, the von Neumann algebra associated with $H$ is
\begin{equation}
\mathcal R(H)=\{W(h):h\in H\}''. \label{eq:RH}
\end{equation}
Let $\Omega$ denote the Fock vacuum and $\omega(A)=\langle\Omega,A\Omega\rangle$. For $\xi\in\Hh$, define the coherent state restricted to $\mathcal R(H)$ by
\begin{equation}
\omega_\xi(A)=\omega\bigl(W(\xi)AW(\xi)^*\bigr). \label{eq:coherentstate}
\end{equation}
The displacement need not itself lie in $H$; only the component visible to $\mathcal R(H)$ contributes to the restricted state. Importantly, the coherent displacement changes the first moment. Its covariance is the vacuum covariance. This is the correct Gaussian-state interpretation used below.

\section{Araki relative entropy}\label{sec:araki}

For faithful normal states in standard form, represented by vectors $\Psi$ and $\Omega$, the relative Tomita--Takesaki operator is defined on the natural core by
\begin{equation}
s_{\Omega,\Psi}(A\Psi)=A^*\Omega. \label{eq:relativetomita}
\end{equation}
Writing its polar decomposition as
\begin{equation}
s_{\Omega,\Psi}=j_{\Omega,\Psi}\,\delta_{\Omega,\Psi}^{1/2}, \label{eq:relativepolar}
\end{equation}
Araki relative entropy \cite{Araki} is
\begin{equation}
S_{\mathcal M}(\Psi\,|\,\Omega)=-\langle\Psi,\log\delta_{\Omega,\Psi}\,\Psi\rangle, \label{eq:arakientropy}
\end{equation}
when the corresponding quadratic form is finite. The invariant zero statement is
\begin{equation}
S_{\mathcal M}(\varphi\,|\,\omega)=0\iff\varphi=\omega\ \text{as normal states on}\ \mathcal M, \label{eq:zero}
\end{equation}
under the usual faithfulness hypotheses.

For coherent states of a CCR algebra the many-body expression reduces to a one-particle quadratic form. The required real-linear operation is the cutting projection.

\subsection{The cutting projection}\label{sec:cutting}

For a factorial standard subspace, vectors in the dense real domain $H+H'$ have a unique decomposition
\begin{equation}
\xi=h+h',\qquad h\in H,\ h'\in H'. \label{eq:decomposition}
\end{equation}
The cutting projection is the closed real-linear operator
\begin{equation}
P_H:\ H+H'\to H,\qquad P_H(h+h')=h. \label{eq:PHdef}
\end{equation}
It is generally unbounded. On its natural domain it admits the modular representation
\begin{equation}
\boxed{\ P_H=(1+s_H)(1-\delta_H)^{-1}.\ } \label{eq:PHmodular}
\end{equation}
The inverse is understood on $\ran(1-\delta_H)$ (or, equivalently, by the appropriate spectral-calculus domain); the possible singularity at spectral value $1$ must not be ignored. Appendix~\ref{app:cutting} gives an elementary algebraic derivation of Eq.~\eqref{eq:PHmodular} on an explicitly stated spectral core; the closed-operator statement on the natural domain is part of the standard-subspace analysis of Ciolli--Longo--Ruzzi \cite{CLR}, who show that the right-hand side has range in $H$, acts as the identity on $H$, and vanishes on $H'$.

\subsection{The one-particle entropy formula}\label{sec:oneparticle}

Following the standard-subspace/coherent-state entropy results of Refs.~\cite{CLR,BCD}, the coherent-state entropy associated with $H$ is the closed quadratic form
\begin{equation}
S_H(\xi)=-\im\langle\xi,\,P_H\,i\log\delta_H\,\xi\rangle, \label{eq:entropyformula}
\end{equation}
with the corresponding form-domain qualification. Its second-quantized meaning is
\begin{equation}
S_{\mathcal R(H)}(\omega_\xi\,|\,\omega)=S_H(\xi). \label{eq:secondquantized}
\end{equation}
Thus no Pauli--Jordan distribution, spacetime test function, or field equation is needed in the abstract formula itself.

If the displacement is localized, $h\in H$, the reduction to a localized quadratic form is not simply the statement $P_Hh=h$: in Eq.~\eqref{eq:entropyformula} the cutting projection acts on $i\log\delta_H\,h$, which is in general not an element of $H$. The correct mechanism is symplectic orthogonality. Writing $k_H:=-\log\delta_H$ and decomposing $i\,k_Hh=a+a'$ with $a=P_H(ik_Hh)\in H$ and $a'\in H'$, the definition~\eqref{eq:sympcompl} gives $\im\langle h,a'\rangle=0$, and hence
\begin{equation}
S_H(h)=-\langle h,\log\delta_H\,h\rangle=\langle h,k_Hh\rangle,\qquad k_H:=-\log\delta_H, \label{eq:localized}
\end{equation}
again in quadratic-form sense. The operator $k_H$ is not positive on the full complex Hilbert space. Positivity of relative entropy is a statement about the appropriate localized closed quadratic form.

\subsection{Spectral representation}\label{sec:spectral}

Let $E_H(\kappa)$ be the spectral measure of $k_H=-\log\delta_H$. Then
\begin{equation}
k_H=\int_{\mathbb R}\kappa\,dE_H(\kappa). \label{eq:spectralkH}
\end{equation}
For a localized $h$ in the form domain,
\begin{equation}
S_H(h)=\int_{\mathbb R}\kappa\,d\mu_h(\kappa),\qquad d\mu_h(\kappa)=d\langle h,E_H(\kappa)h\rangle. \label{eq:spectralentropy}
\end{equation}
The measure $d\mu_h$ is not arbitrary: the reality condition
\begin{equation}
s_Hh=h\iff j_H\delta_H^{1/2}h=h \label{eq:reality}
\end{equation}
relates the positive and negative modular spectral components. For the wedge subspace this relation will be made completely explicit below.

\section{Rindler wedge and the absence of explicit mass in one isolated modular pair}\label{sec:rindler}

For the right Rindler wedge of a massive scalar field in $1{+}1$ dimensions, use rapidity $\theta$,
\begin{equation}
p_m(\theta)=m(\cosh\theta,\sinh\theta),\qquad \Hh_m\simeq L^2(\mathbb R,d\theta). \label{eq:rapidity}
\end{equation}
A Lorentz boost acts as a rapidity translation. By Bisognano--Wichmann \cite{BW}, we fix the convention
\begin{equation}
(\delta_{\W,m}^{it}\psi)(\theta)=\psi(\theta-2\pi t). \label{eq:BWconvention}
\end{equation}
If $D=-i\partial_\theta$ is the self-adjoint generator of rapidity translations, this means
\begin{equation}
\delta_{\W,m}^{it}=e^{-i2\pi tD},\qquad \log\delta_{\W,m}=-2\pi D,\qquad k_\W=+2\pi D, \label{eq:kW}
\end{equation}
and the modular conjugation acts on rapidity wave functions of the neutral scalar as complex conjugation,
\begin{equation}
(j_\W\psi)(\theta)=\overline{\psi(\theta)}. \label{eq:jW}
\end{equation}
This convention is used consistently below; reversing the boost orientation reverses the corresponding generator signs and the strip orientation used below, but not the entropy statements. Consequently, after rapidity identification the wedge modular generator contains no explicit scalar mass. The abstract wedge standard subspaces for different masses are correspondingly unitarily equivalent when only this isolated modular pair is retained.

This has an immediate consequence. If one chooses the \emph{same abstract modular vector $h(\theta)$} for every mass, then the value of Eq.~\eqref{eq:localized} is the same for every mass. A non-trivial mass curve cannot come from $\delta_\W$ alone.

For later use we record the classical rapidity-space description of the wedge subspace itself. Let $S_\pi:=\{\theta\in\mathbb C:\,0<\im\theta<\pi\}$ and let $H^2(S_\pi)$ denote the Hardy space of functions analytic in $S_\pi$ with $\sup_{0<y<\pi}\int_{\mathbb R}|\psi(\theta+iy)|^2\,d\theta<\infty$; such functions possess $L^2$ boundary values on both boundary lines.

The wedge subspace can now be characterized directly in rapidity space. With the conventions \eqref{eq:BWconvention}--\eqref{eq:jW},
\begin{equation}
H_m(\W_R)=\Bigl\{\psi\in L^2(\mathbb R,d\theta):\ \psi\in H^2(S_\pi)\ \text{and}\ \psi(\theta+i\pi)=\overline{\psi(\theta)}\ \text{for a.e.\ }\theta\in\mathbb R\Bigr\}. \label{eq:stripchar}
\end{equation}
With $\widehat\psi(k)=(2\pi)^{-1/2}\int\psi(\theta)e^{-ik\theta}d\theta$, Eq.~\eqref{eq:kW} gives $\delta_{\W,m}^{1/2}=e^{-\pi D}$, which acts as multiplication by $e^{-\pi k}$ on $\widehat\psi$. By the Paley--Wiener theorem for the strip, $\dom\delta^{1/2}_{\W,m}=\{\psi:\ e^{-\pi k}\widehat\psi\in L^2\}=H^2(S_\pi)$, and on this domain $(\delta^{1/2}_{\W,m}\psi)(\theta)=\psi(\theta+i\pi)$ in the sense of $L^2$ boundary values. By Eq.~\eqref{eq:recover}, $H_m(\W_R)$ is the fixed-point set of $s_\W=j_\W\delta^{1/2}_{\W,m}$, i.e.\ $\overline{\psi(\theta+i\pi)}=\psi(\theta)$.

\subsection{From the Pauli--Jordan representation to modular rapidity}\label{sec:bridge}

The preceding observation raises a natural question. In a spacetime formulation the scalar mass is manifest already at the kinematical level through the Pauli--Jordan distribution, whereas after the natural rapidity identification the modular generator of one isolated Rindler wedge contains no explicit mass parameter. These statements are not in tension: the mass dependence has not disappeared, but has been transferred from the mass-independent isolated-wedge generator to the map that embeds a fixed spacetime preparation into the one-particle Hilbert space.

For a real spacetime test function $f$, let $\Delta_m$ denote the Pauli--Jordan distribution and write the corresponding symplectic form schematically as
\begin{equation}
\sigma_m(f,g)=\int d^dx\,d^dy\ f(x)\,\Delta_m(x-y)\,g(y). \label{eq:PJ}
\end{equation}
Here $d$ denotes the spacetime dimension, so that $\mathbf x,\mathbf y,\mathbf k\in\mathbb R^{d-1}$ (the later notation $1+d$ instead counts $d$ spatial dimensions). With the Fourier convention below and the standard positive-energy measure $d^{d-1}\mathbf k/[(2\pi)^{d-1}2\omega_{\mathbf k}]$, the real Pauli--Jordan distribution is
\begin{equation}
\Delta_m(x-y)=\int_{\mathbb R^{d-1}}\frac{d^{d-1}\mathbf k}{(2\pi)^{d-1}}\,
\frac{\sin\!\bigl[\omega_{\mathbf k}(x^0-y^0)-\mathbf k\cdot(\mathbf x-\mathbf y)\bigr]}{\omega_{\mathbf k}},
\qquad \omega_{\mathbf k}=\sqrt{|\mathbf k|^2+m^2}.
\label{eq:PJmomentum}
\end{equation}
The integral is understood distributionally. The positive sign and the factor $1/\omega_{\mathbf k}$ follow from $2\im e^{ip\cdot(x-y)}=2\sin[p\cdot(x-y)]$: with the inner product antilinear in its first argument, this is precisely the kernel of $2\im\langle E_mf,E_mg\rangle$ in Eq.~\eqref{eq:sigmaIm}.
The dependence on $m$ is explicit because $\Delta_m$ restricts Fourier data to the mass shell. In $1{+}1$ dimensions, on the positive-energy shell
\begin{equation}
p_m(\theta)=m(\cosh\theta,\sinh\theta),\qquad \frac{dp^1}{2p^0}=\frac{d\theta}{2}. \label{eq:massshell}
\end{equation}
Define the mass-shell restriction map\footnote{With the Fourier convention $\widehat f(p^0,p^1)=\int_{\mathbb R^2}dt\,dx\;e^{-i(p^0t-p^1x)}f(t,x)$, the rapidity representative is explicitly
\[
h_m(\theta)=c\int_{\mathbb R^2}dt\,dx\;e^{-i(m\cosh\theta\,t-m\sinh\theta\,x)}f(t,x).
\]
For a right-wedge-localized preparation one takes $f\in C_c^\infty(\W_R)$. The sign of the exponent is tied to the strip orientation described above: for real $f$ supported in $\W_R$ the continued exponent has real part $-m\sin y\,(x\cosh\theta-t\sinh\theta)<0$ throughout $0<y=\im\theta<\pi$, and $h_m(\theta+i\pi)=\overline{h_m(\theta)}$, consistently with Eq.~\eqref{eq:stripchar}.}
\begin{equation}
E_m:\ f\longmapsto h_m,\qquad (E_mf)(\theta)=h_m(\theta)=c\,\widehat f\bigl(m\cosh\theta,\,m\sinh\theta\bigr), \label{eq:Em}
\end{equation}
where $c=(4\pi)^{-1/2}$ for the measure specified above and the rapidity inner product $\langle h_1,h_2\rangle=\int d\theta\,\overline{h_1(\theta)}h_2(\theta)$. With these conventions,
\begin{equation}
\sigma_m(f,g)=2\,\im\langle E_mf,E_mg\rangle. \label{eq:sigmaIm}
\end{equation}
Thus the same mass-shell datum that is visible in $\Delta_m$ is encoded, in the one-particle description, in $E_m$.

For a wedge-localized coherent excitation generated by the fixed spacetime preparation $f$, the one-particle Araki relative entropy therefore becomes
\begin{equation}
S_m[f]=\langle E_mf,\ k_\W\,E_mf\rangle,\qquad k_\W=-\log\delta_\W, \label{eq:Smf}
\end{equation}
in the appropriate quadratic-form domain. Bisognano--Wichmann makes $k_\W$ mass-independent after rapidity identification, but the vector on which it acts generally depends on the mass:
\begin{equation}
h_m(\theta)=c\,\widehat f(m\cosh\theta,m\sinh\theta)\,. \label{eq:hm}
\end{equation}
Consequently
\begin{equation}
\partial_mk_\W=0\quad\text{does not imply}\quad \partial_mS_m[f]=0. \label{eq:noimplication}
\end{equation}
When $m\mapsto E_mf$ is differentiable in the relevant form topology,
\begin{equation}
\frac{dS_m[f]}{dm}=2\,\re\bigl\langle(\partial_mE_m)f,\ k_\W\,E_mf\bigr\rangle. \label{eq:derivative}
\end{equation}
No universal sign follows from this identity alone: it depends on the spacetime preparation held fixed in the comparison.

This resolves the apparent mismatch between the two formulations. Holding the same abstract rapidity vector $h(\theta)$ fixed for every $m$ removes the mass dependence by construction. Holding instead the same spacetime test function $f(x)$ fixed produces the family $h_m=E_mf$ and generically a nontrivial mass curve. The two prescriptions compare different physical data.

The same conclusion is intrinsic to modular localization. Translations act as
\begin{equation}
(U_m(a)h)(\theta)=e^{ip_m(\theta)\cdot a}h(\theta), \label{eq:translations}
\end{equation}
and in null coordinates
\begin{equation}
P_+=\frac{m}{\sqrt2}\,e^\theta,\qquad P_-=\frac{m}{\sqrt2}\,e^{-\theta},\qquad 2P_+P_-=m^2. \label{eq:nullcoords}
\end{equation}
Hence the mass is retained by the translation representation and by the translated family $a\mapsto H_m(\W+a)$. In concise form,
\begin{equation}
\Delta_m\ \longleftrightarrow\ E_m\ \longleftrightarrow\ h_m=E_mf\ \longleftrightarrow\ S_m[f], \label{eq:chain}
\end{equation}
while the isolated wedge generator remains mass-independent after rapidity identification.

For a general displacement $\xi_m$ that is not already wedge localized, the bridge is completed by the same cutting projection used throughout this paper,
\begin{equation}
f\ \longmapsto\ \xi_m=E_mf\ \longmapsto\ P_{H_m(\W)}\xi_m\ \longmapsto\ S_{H_m(\W)}(\xi_m), \label{eq:bridgechain}
\end{equation}
with
\begin{equation}
P_{H_m(\W)}=(1+s_{H_m(\W)})(1-\delta_\W)^{-1} \label{eq:PHm}
\end{equation}
on its natural real domain. Appendix~\ref{app:cutting} gives the detailed cutting-projector derivation. The present section supplies the complementary physical bridge: the mass dependence that is explicit in the Pauli--Jordan kernel is carried into modular theory by the mass-shell embedding and, equivalently, by the mass-dependent translation/localization data.

In $1+d$ dimensions the same construction uses the transverse mass
\begin{equation}
\mu_m(p_\perp)=\sqrt{m^2+|p_\perp|^2}, \label{eq:transversemass}
\end{equation}
so that
\begin{equation}
h_m(\theta,p_\perp)=c\,\widehat f\bigl(\mu_m(p_\perp)\cosh\theta,\ \mu_m(p_\perp)\sinh\theta,\ p_\perp\bigr)\,. \label{eq:hmhigher}
\end{equation}
This gives a constructive route to higher-dimensional mass curves without assigning the mass to the isolated wedge modular Hamiltonian itself.

\subsection{How the mass re-enters}\label{sec:reentry}

The mass has not disappeared from the physical theory. It re-enters through the translation representation and through the rule used to compare excitations in different theories. In null coordinates in $1{+}1$ dimensions, Eq.~\eqref{eq:nullcoords} shows that the joint translation spectrum recovers the mass-shell invariant even though the boost action alone does not. This also makes clear why one should be cautious with a single half-sided/null inclusion: one null generator can be rescaled, whereas retaining sufficient translation data---for example both null generators and their joint relation---makes the mass invariant visible. Related modular analyses of massless and null-plane settings further illustrate the special role of null localization \cite{LM,MTW}. Two modular descriptions are useful.

\subsubsection{A mass-dependent family of coherent vectors}\label{sec:family}

Choose
\begin{equation}
m\longmapsto h_m\in H_m(\W_R). \label{eq:familychoice}
\end{equation}
Then
\begin{equation}
S(m)=S_{H_m(\W_R)}(h_m)=-\langle h_m,\log\delta_{\W,m}\,h_m\rangle. \label{eq:familyentropy}
\end{equation}
Even when the modular operator becomes mass-independent after rapidity identification, the vector $h_m$ need not.

\subsection{The translated modular net}\label{sec:net}

Retain the family of translated standard subspaces
\begin{equation}
a\longmapsto H_m(\W_R+a). \label{eq:translatednet}
\end{equation}
Translations act in rapidity as in Eq.~\eqref{eq:translations}, so that the mass appears explicitly through $p_m$. Hence the full localization structure distinguishes masses even though one isolated wedge modular operator does not. In this sense, mass is encoded in the relative position of standard subspaces under translations, rather than in the boost modular operator of one isolated wedge.

\section{Four conditions and an entire localized \texorpdfstring{$1+1$}{1+1} construction}\label{sec:fourconditions}

The bridge $f\mapsto h_m=E_mf$ identifies the correct place at which mass enters the one-particle problem. More precisely, the mass-shell structure induced by this map yields four structural conditions on a rapidity representative:

\begin{description}
\item[Condition 1.] \emph{On-shell functional dependence.}
\begin{equation}
\left(m\frac{\partial}{\partial m}-\cosh\theta\,\frac{\partial}{\partial(\cosh\theta)}-\sinh\theta\,\frac{\partial}{\partial(\sinh\theta)}\right)h(m,\theta)=0. \label{eq:condition1}
\end{equation}
Its characteristic variables are
\begin{equation}
p^0=m\cosh\theta,\qquad p^1=m\sinh\theta,\qquad p_\pm=me^{\pm\theta}. \label{eq:characteristics}
\end{equation}
Equivalently, $h(m,\theta)=F\bigl(me^{\theta},\,me^{-\theta}\bigr)$ for a single function $F$ of the on-shell light-cone momenta. Condition~1 is the infinitesimal form of an on-shell change of variables: it fixes the functional dependence of the family on $(m,\theta)$ and is a bookkeeping constraint, not a dynamical selection principle.
\item[Condition 2.] \emph{Controlled massless boundary value.}
\begin{equation}
\lim_{m\to0^+}h(m,\theta)=C,\qquad C\ \text{independent of}\ \theta. \label{eq:condition2}
\end{equation}
For the family constructed below $C=0$, and the nontrivial massless content appears in a comoving rapidity frame; see Remark~\ref{rem:masslesslimit}.
\item[Condition 3.] \emph{Decay at large real rapidity.}
\begin{equation}
h(m,\theta)\longrightarrow0\qquad(\theta\to\pm\infty,\ m>0). \label{eq:condition3}
\end{equation}
\item[Condition 4.] \emph{Bisognano--Wichmann strip analyticity.}
\begin{equation}
h(m,\theta)\ \text{is analytic for}\qquad 0<\im\theta<\pi. \label{eq:condition4}
\end{equation}
\end{description}

These conditions make clear that the mass dependence is a property of the vector family rather than of the isolated boost generator. They are organizing conditions; the sharp membership criterion is the rapidity characterization above, which in addition to Condition~4 requires the Hardy-type $L^2$ control in the strip and the Tomita boundary relation. The family constructed below satisfies all of them.

\begin{remark}[Why sinc factors, a doubled phase, and a Gaussian pair]\label{rem:motivation}
It is instructive to see how a simpler ansatz fails. The rational-phase family
$e^{im\sinh\theta}\bigl[(1+me^{\theta})(1+me^{-\theta})\bigr]^{-1}$
satisfies Conditions 1--4 pointwise, but its continuation to the upper boundary line,
$e^{-im\sinh\theta}\bigl[(1-me^{\theta})(1-me^{-\theta})\bigr]^{-1}$,
has non-integrable singularities at $\theta=\mp\log m$ and does not equal the complex conjugate of the function on the real line; it therefore fails both requirements of the rapidity characterization and can serve only as an unlocalized seed. The construction below repairs exactly these two failures: entire sinc factors remove the boundary poles, the \emph{doubled} light-cone phase produces an interior-strip decay that cancels the sinc growth exactly, and the Gaussian \emph{pair} restores real-axis decay while transforming correctly under the strip shift $\theta\mapsto\theta+i\pi$.
\end{remark}

The explicit family is
\begin{align}
h_m(\theta)
&=\bigl(1+s_{H_m}\bigr)\left[
\frac{1}{a}\,e^{2im\sinh\theta}\,
\sinc(me^\theta)\,\sinc(me^{-\theta})
e^{-(\theta+\log m)^2}
\right] \notag\\
&=\frac{1}{a}\,e^{2im\sinh\theta}\,
\sinc(me^\theta)\,\sinc(me^{-\theta})
\Bigl[e^{-(\theta+\log m-i\pi)^2}+e^{-(\theta+\log m)^2}\Bigr],
\label{eq:entirefamily}
\end{align}
where $\sinc z:=\sin z/z$ with its removable value at $z=0$, and $a>0$ is a dimensionless normalization constant. Throughout, masses are measured in units of a fixed reference mass $M>0$; that is, $m$ and $p_\pm$ in Eq.~\eqref{eq:entirefamily} and below denote the dimensionless ratios $m/M$ and $p_\pm/M$, so that $\log m$ and the sinc and phase arguments are well defined. Write
\begin{equation}
p_+=me^\theta,\qquad p_-=me^{-\theta},\qquad
2m\sinh\theta=p_+-p_-,\qquad \theta+\log m=\log p_+.
\label{eq:lightconeidentities}
\end{equation}
Then
\begin{equation}
h_m(\theta)=\frac{1}{a}\,e^{i(p_+-p_-)}\,
\sinc(p_+)\,\sinc(p_-)
\Bigl[e^{-(\log p_+-i\pi)^2}+e^{-(\log p_+)^2}\Bigr].
\label{eq:lightconeform}
\end{equation}
Thus all mass and rapidity dependence occurs through the on-shell light-cone variables.

\subsection{Differential constraint}\label{sec:diffconstraint}

Since $h_m(\theta)=F(p_+,p_-)$, it obeys identically the characteristic equation of Eq.~\eqref{eq:condition1},
\begin{equation}
\bigl(m\partial_m-p_+\partial_{p_+}-p_-\partial_{p_-}\bigr)h_m=0.
\label{eq:diffconstraint}
\end{equation}
This is the precise sense in which the family satisfies the mass-shell differential condition.

\subsection{Massless and large-mass behavior; real-rapidity decay}\label{sec:masslessbehavior}

For fixed real $\theta$,
\begin{equation}
\lim_{m\to0^+}h_m(\theta)=0,
\label{eq:masslesslimit}
\end{equation}
independently of $\theta$. This is only a pointwise statement: the Gaussian packet translates towards $\theta\simeq-\log m$, and in the comoving coordinate $q=\theta+\log m$ a nontrivial profile remains. Strong $L^2$ convergence as $m\to0^+$ is therefore not asserted; see Remark~\ref{rem:masslesslimit} for the precise statement. For fixed $\theta$, the Gaussian displacement and the sinc suppression also imply $h_m(\theta)\to0$ as $m\to\infty$.

For real $\theta$ the phase has unit modulus, the sinc factors are bounded, and the Gaussian pair has Gaussian falloff. Hence $h_m(\theta)\to0$ as $\theta\to\pm\infty$, $h_m\in L^2(\mathbb R,d\theta)$ for every $m>0$, and, since every $\theta$-derivative of $h_m$ again has Gaussian decay on the real axis, $h_m$ lies in the domain of $D$ and in the form domain of $k_\W$.

\subsection{Entire analyticity, Hardy control, and the Tomita--Takesaki boundary relation}\label{sec:analyticity}

Every factor in Eq.~\eqref{eq:entirefamily} is entire in $\theta$; the apparent singularities of the sinc functions are removable. Moreover,
\begin{equation}
\sinh(\theta+i\pi)=-\sinh\theta,\qquad e^{\pm(\theta+i\pi)}=-e^{\pm\theta}.
\label{eq:stripshift}
\end{equation}
Because $\sinc(-z)=\sinc(z)$, the doubled phase is conjugated by the strip shift. With $x=\theta+\log m$ and
\begin{equation}
G_m(\theta)=e^{-(x-i\pi)^2}+e^{-x^2},
\label{eq:gaussianpair}
\end{equation}
one also has $G_m(\theta+i\pi)=\overline{G_m(\theta)}$ for real $\theta$. Consequently,
\begin{equation}
h_m(\theta+i\pi)=\overline{h_m(\theta)}.
\label{eq:tomitaboundary}
\end{equation}
This is exactly the Tomita boundary relation used in the rapidity characterization above.

The doubled phase supplies the required interior estimate. For $z=x'+iy$ with $0<y<\pi$, put $u=me^{x'}$ and $v=me^{-x'}$. Then
\begin{equation}
\bigl|e^{2im\sinh z}\bigr|=e^{-(u+v)\sin y},\qquad
\bigl|\sinc(me^{z})\bigr|\le C\,\frac{e^{u\sin y}}{1+u},\qquad
\bigl|\sinc(me^{-z})\bigr|\le C\,\frac{e^{v\sin y}}{1+v}.
\label{eq:stripcancellation}
\end{equation}
The exponential factors cancel exactly, while
\begin{equation}
|G_m(x'+iy)|\le 2e^{\pi^2}e^{-(x'+\log m)^2},\qquad 0\le y\le\pi.
\label{eq:gaussianbound}
\end{equation}
It follows that
\begin{equation}
\sup_{0<y<\pi}\int_{-\infty}^{\infty}|h_m(x'+iy)|^2\,dx'<\infty.
\label{eq:hardybound}
\end{equation}
Thus $h_m$ is entire, with the required Hardy-type $L^2$ control throughout the whole Bisognano--Wichmann strip. (In the lower strip $-\pi<y<0$ the phase and the sinc factors grow \emph{jointly} like $e^{2(u+v)|\sin y|}$; the construction is intrinsically tied to the strip orientation fixed in Section~\ref{sec:rindler}.)

\subsection{Localization of the family}\label{sec:localizationtheorem}

We can now verify the wedge localization of the family: for every $m>0$, the vector $h_m$ of Eq.~\eqref{eq:entirefamily} belongs to $H_m(\W_R)$.

By Section~\ref{sec:masslessbehavior}, $h_m\in L^2(\mathbb R,d\theta)$. By Section~\ref{sec:analyticity}, $h_m$ is entire, satisfies the Hardy bound \eqref{eq:hardybound}---so its restriction to $S_\pi$ lies in $H^2(S_\pi)$, with boundary values attained continuously---and obeys the boundary relation \eqref{eq:tomitaboundary}. These are precisely the conditions stated in the rapidity characterization above.

Together with the rapidity characterization above and the established one-particle formula \eqref{eq:secondquantized}--\eqref{eq:localized}, this shows that the quantity computed in the next subsection is the Araki relative entropy of the coherent state $\omega_{h_m}$ with respect to the vacuum on $\mathcal R(H_m(\W_R))$---not merely a candidate.

\subsection{The entropy: positivity, spectral form, and the mass curve}\label{sec:entropycurve}

With the convention \eqref{eq:kW}, the localized entropy \eqref{eq:localized} of the family is
\begin{equation}
S(m)\;=\;\langle h_m,\,k_\W\,h_m\rangle\;=\;2\pi\,\langle h_m,\,D\,h_m\rangle
\;=\;-2\pi\,\re\int_{-\infty}^{\infty}\overline{h_m(\theta)}\,i\partial_\theta h_m(\theta)\,d\theta.
\label{eq:entropydef}
\end{equation}
The real-axis Gaussian falloff of $h_m$ and of its derivative makes the integral absolutely convergent for every $m>0$. The normalization constant only rescales the answer:
\begin{equation}
S(m;a)=a^{-2}\,S(m;1).
\label{eq:normalizationscaling}
\end{equation}

Positivity is not a convention here; it is a consequence of the boundary relation.

The boundary relation also gives a direct spectral representation of the entropy. Let $\psi\in H_m(\W_R)$ lie in the form domain of $k_\W$, and let $\widehat\psi(k)=(2\pi)^{-1/2}\int\psi(\theta)e^{-ik\theta}d\theta$. It implies
\begin{equation}
e^{-\pi k}\,\widehat\psi(k)=\overline{\widehat\psi(-k)},\qquad\text{hence}\qquad
|\widehat\psi(-k)|^2=e^{-2\pi k}\,|\widehat\psi(k)|^2 ,
\label{eq:FTboundary}
\end{equation}
and
\begin{equation}
S_{H_m(\W_R)}(\psi)=2\pi\int_0^\infty k\,\bigl|\widehat\psi(k)\bigr|^2\bigl(1-e^{-2\pi k}\bigr)\,dk\;\ge\;0,
\label{eq:spectralpositivity}
\end{equation}
with equality if and only if $\psi=0$.
Taking Fourier transforms of $\psi(\cdot+i\pi)=\overline{\psi(\cdot)}$: the left side is $e^{-\pi k}\widehat\psi(k)$ (contour shift, justified by the Hardy bound), the right side is $\overline{\widehat\psi(-k)}$. Then
$\langle\psi,D\psi\rangle=\int_{\mathbb R}k\,|\widehat\psi(k)|^2dk
=\int_0^\infty k\bigl(|\widehat\psi(k)|^2-|\widehat\psi(-k)|^2\bigr)dk$,
and Eq.~\eqref{eq:FTboundary} gives \eqref{eq:spectralpositivity}. If the integral vanishes, $\widehat\psi=0$ a.e.\ on $(0,\infty)$, hence by \eqref{eq:FTboundary} also on $(-\infty,0)$, so $\psi=0$.

Equation~\eqref{eq:spectralpositivity} is the completely explicit form, for the wedge, of the spectral reality constraint of Section~\ref{sec:spectral}: the negative modular frequencies of a localized vector are exponentially tied to the positive ones, and the entropy is a manifestly positive integral over the positive modular spectrum. It also provides a strong independent check on the numerics below: evaluating \eqref{eq:entropydef} directly and evaluating \eqref{eq:spectralpositivity} via the Fourier transform of $h_m$ agree to $15$ significant digits at $m=0.1,\,1,\,3$.

\begin{remark}[The massless limit runs along the modular flow]\label{rem:masslesslimit}
The escape of the packet to $\theta\simeq-\log m$ is a rapidity translation, i.e.\ precisely the modular flow \eqref{eq:BWconvention}, under which the quadratic form \eqref{eq:entropydef} is invariant. In the comoving variable $q=\theta+\log m$ the family converges pointwise, as $m\to0^+$, to the profile
$h_0(q)=a^{-1}e^{ie^{q}}\sinc(e^{q})\bigl[e^{-(q-i\pi)^2}+e^{-q^2}\bigr]$,
which itself satisfies the Tomita boundary relation, and numerically
$S(m)\to S[h_0]\approx1.5299\,S(1)$.
This explains why the pointwise limit \eqref{eq:masslesslimit} is $0$ while the entropy has a finite massless limit: the vector does not go to zero in $L^2$; it runs away along the modular flow.
\end{remark}

\begin{figure}[t]
\centering
\includegraphics[width=0.74\textwidth]{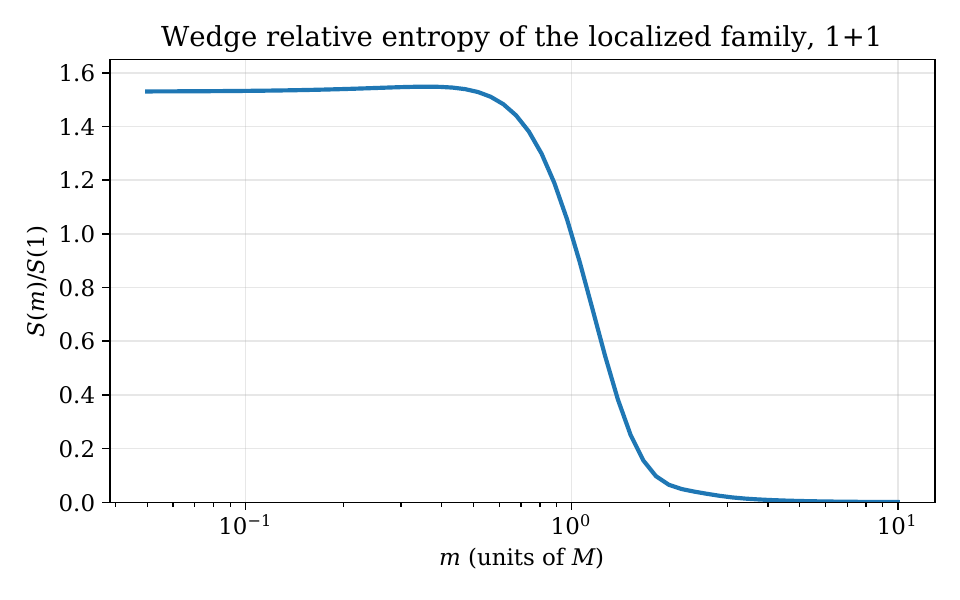}
\caption{Mass dependence of the Araki relative entropy $S(m)/S(1)$ of the localized family \eqref{eq:entirefamily} in $1{+}1$ dimensions, Eq.~\eqref{eq:entropydef}, with $m$ in units of the reference mass $M$ and $a$ held fixed (ratios are $a$-independent). The massless plateau equals the modular-flow limit of Remark~\ref{rem:masslesslimit}; the maximum is at $m\approx0.37$; the large-mass regime is strongly suppressed.}
\label{fig:masscurve}
\end{figure}

Direct numerical evaluation of Eq.~\eqref{eq:entropydef} for the complete complex wave function---including the derivative of the doubled phase, both sinc factors, and the Gaussian pair---gives, at $a=1$, the absolute scale $S(1)\approx8.884\times10^{9}$ (the large number simply reflects the unnormalized factor $e^{\pi^2}\approx1.93\times10^{4}$ carried by the first Gaussian; only ratios are shown below) and
\begin{center}
\begin{tabular}{c|c}
$m$ & $S(m)/S(1)$\\ \hline
$0.05$ & $1.53059$\\
$0.10$ & $1.53261$\\
$0.30$ & $1.54659$\\
$1.00$ & $1.00000$\\
$3.00$ & $0.0205826$\\
$10.0$ & $0.00104230$
\end{tabular}
\end{center}
The curve is not monotone: it rises mildly from its massless limit to a maximum $S/S(1)\approx1.548$ at $m\approx0.37$, and is then strictly decreasing, with strong suppression at large mass (the local logarithmic slope is $\approx-2.2$ over $5\le m\le 20$). The displayed profile is a property of this explicit mass-indexed family, not a universal mass dependence. The maximum is therefore stated in units of the reference scale $M$ built into the family. The quoted slope characterizes only the displayed finite range, not an asserted asymptotic exponent. See Fig.~\ref{fig:masscurve}.

\section{Extension to \texorpdfstring{$1+d$}{1+d} dimensions}\label{sec:higherd}

The extension follows the transverse-mass decomposition of Section~\ref{sec:bridge}. For fixed transverse momentum $p_\perp$, define
\begin{equation}
\mu_m(p_\perp)=\sqrt{m^2+|p_\perp|^2}.
\label{eq:transversemass2}
\end{equation}
Every transverse fiber is a copy of the $1+1$ construction, with $m$ replaced by $\mu=\mu_m(p_\perp)$:
\begin{equation}
h_m^{(d)}(\theta,p_\perp)=\frac{\chi(p_\perp)}{a}\,e^{2i\mu\sinh\theta}\,
\sinc(\mu e^\theta)\,\sinc(\mu e^{-\theta})
\Bigl[e^{-(\theta+\log\mu-i\pi)^2}+e^{-(\theta+\log\mu)^2}\Bigr].
\label{eq:higherfamily}
\end{equation}
Here $\chi$ is a fixed normalized transverse wave packet, independent of $m$; no additional mass-dependent ansatz is introduced. In higher dimensions the modular conjugation acts on the transverse multiplicity as
\begin{equation}
(j_\W\psi)(\theta,p_\perp)=\overline{\psi(\theta,-p_\perp)},
\label{eq:jWhigher}
\end{equation}
so the rapidity characterization holds fiberwise with the boundary relation $\psi(\theta+i\pi,p_\perp)=\overline{\psi(\theta,-p_\perp)}$. Membership of \eqref{eq:higherfamily} therefore requires a reality condition on the packet itself, not only on $|\chi|^2$:
\begin{equation}
\chi(-p_\perp)=\overline{\chi(p_\perp)}.
\label{eq:chireality}
\end{equation}
We take the real, even, normalized Gaussians
\begin{equation}
\chi(p)=\pi^{-1/4}e^{-p^2/2}\ \ (1{+}2),\qquad
\chi(p_\perp)=\pi^{-1/2}e^{-|p_\perp|^2/2}\ \ (1{+}3),
\label{eq:gaussianpackets}
\end{equation}
which satisfy \eqref{eq:chireality}.

The same construction extends directly to $1+d$ dimensions. With $\chi$ as in \eqref{eq:chireality}--\eqref{eq:gaussianpackets}, $h^{(d)}_m\in H_m(\W_R)$ for every $m>0$, and
\begin{equation}
S^{1+d}(m)=\int_{\mathbb R^{d-1}}d^{\,d-1}p_\perp\,
|\chi(p_\perp)|^2\,f\!\bigl(\mu_m(p_\perp)\bigr),
\qquad
f(\mu):=-2\pi\,\re\int_{-\infty}^{\infty}\overline{h_\mu(\theta)}\,i\partial_\theta h_\mu(\theta)\,d\theta,
\label{eq:transverseaverage}
\end{equation}
is finite and strictly positive, where $h_\mu$ is the $1{+}1$ family \eqref{eq:entirefamily} at mass $\mu$.
For each fixed $p_\perp$ the fiber satisfies, verbatim, the entirety, Hardy bound, and $+i\pi$ boundary relation of Section~\ref{sec:analyticity} with $m\to\mu$; the fiber Hardy constant is uniform in $\mu$ because the Gaussian bound \eqref{eq:gaussianbound} is translation invariant in $x'$. Condition \eqref{eq:chireality} converts the fiberwise relation into the full boundary relation for \eqref{eq:jWhigher}, and dominated convergence gives the vector-valued Hardy property; hence membership follows from the fiberwise rapidity characterization above. Since $k_\W$ acts on rapidity only, the quadratic form factorizes as \eqref{eq:transverseaverage}; each $f(\mu)>0$ by the spectral representation above, and the Gaussian weight makes the transverse average finite.

Numerically, with the packets \eqref{eq:gaussianpackets} and each curve normalized at $m=1$:
\begin{center}
\begin{tabular}{c|ccc}
$m$ & $S^{1+1}(m)/S^{1+1}(1)$ & $S^{1+2}(m)/S^{1+2}(1)$ & $S^{1+3}(m)/S^{1+3}(1)$\\ \hline
$0.05$ & $1.53059$ & $1.87318$ & $2.15113$\\
$0.10$ & $1.53261$ & $1.87064$ & $2.14422$\\
$0.30$ & $1.54659$ & $1.83670$ & $2.06571$\\
$1.00$ & $1.00000$ & $1.00000$ & $1.00000$\\
$3.00$ & $0.02058$ & $0.02691$ & $0.03560$\\
$10.0$ & $0.00104$ & $0.00146$ & $0.00207$
\end{tabular}
\end{center}

\begin{figure}[t]
\centering
\includegraphics[width=0.74\textwidth]{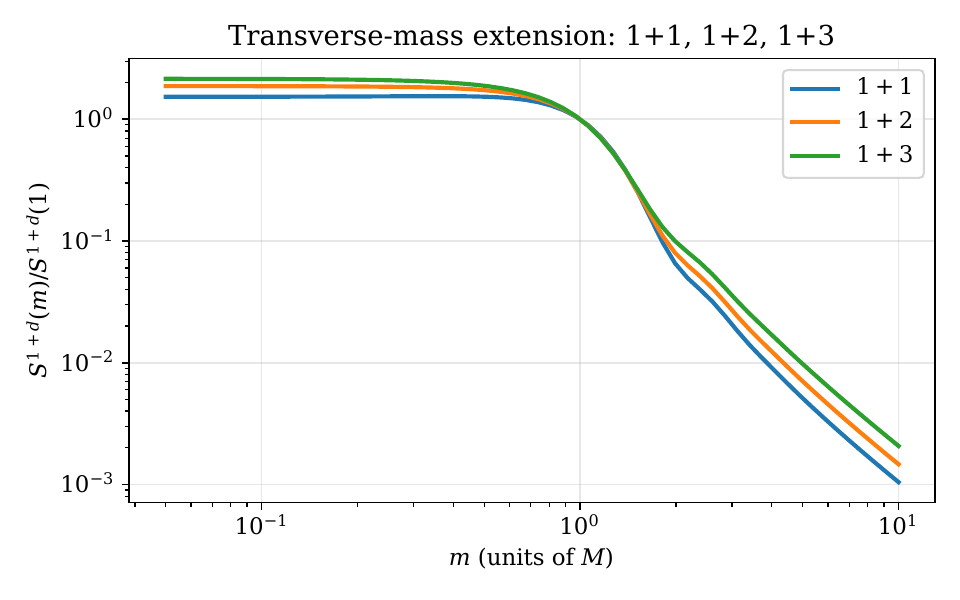}
\caption{Araki relative entropy of the localized family in $1{+}1$, $1{+}2$ and $1{+}3$ dimensions, normalized separately at $m=1$ (units and normalization as in Fig.~\ref{fig:masscurve}). The higher-dimensional curves are Gaussian transverse averages, Eq.~\eqref{eq:transverseaverage}, of the same fiber function; their positivity and finiteness follow from the discussion above.}
\label{fig:highercurves}
\end{figure}

\subsection{Large-mass behavior}\label{sec:largemass}

The two sinc factors suppress configurations for which either light-cone variable becomes large, and the Gaussian packet is displaced with $\log\mu$. Since $p_+p_-=\mu^2$, increasing $\mu$ forces at least one sinc argument away from the small-argument region throughout rapidity space. This yields the large-mass suppression seen in the $1+1$ curve; dominated transverse averaging transfers the suppression to $1+2$ and $1+3$ dimensions.

\section{Conclusions}\label{sec:conclusions}

The modular Hamiltonian of the isolated right wedge is mass-independent after the unitary rapidity identification, but the physical mass enters through the mass-shell embedding of the one-particle vectors on which that operator is evaluated. This observation resolves the apparent tension between the mass independence of the Bisognano--Wichmann generator and the mass dependence of coherent-state relative entropy.

The central constructive result is the explicit entire family \eqref{eq:entirefamily} and its localization. The doubled light-cone phase cancels the interior-strip growth of the two sinc factors exactly; the Gaussian pair supplies real-axis decay and transforms correctly under $\theta\mapsto\theta+i\pi$. The family depends only on the on-shell variables, satisfies the Tomita boundary relation, and has uniform Hardy-type $L^2$ control in the Bisognano--Wichmann strip; the rapidity characterization of the wedge subspace then gives $h_m\in H_m(\W_R)$ for every $m>0$. Its Araki relative entropy is therefore an unconditional, finite, strictly positive quantity, with the exact spectral representation \eqref{eq:spectralpositivity} in which positivity is manifest. The computed mass curve has a finite massless limit attained along the modular flow (Remark~\ref{rem:masslesslimit}), a maximum at intermediate mass $m\approx0.37$ in $1+1$ dimension, and strong large-mass suppression.

Section~\ref{sec:higherd} shows that no new mass-dependent ansatz is needed to pass to higher dimension. The $1+d$ construction is obtained fiber by fiber by the single replacement
\[
m\ \longmapsto\ \mu_m(p_\perp)=\sqrt{m^2+|p_\perp|^2},
\]
followed by the transverse average with a fixed normalized wave packet obeying the reality condition \eqref{eq:chireality}. In this precise sense, the higher-dimensional mass dependence is inherited from the explicit modular one-particle construction rather than imposed through an additional phenomenological damping law. The remaining freedom is the choice of the transverse state itself: a genuine choice of one-particle vector, not an extra prescription for how the mass should enter.

Several natural questions remain open. First, the family \eqref{eq:entirefamily} is specified intrinsically in rapidity space; identifying an explicit spacetime preparation class realizing it through $E_m$ (the sinc factors correspond to sharp light-cone profiles, while the Gaussian in $\log p_+$ is not of compact-support Paley--Wiener type) would close the circle with the spacetime-smearing computations of Refs.~\cite{GRSV,CGRS,Predecessor}. Second, the massless limit exists only along the modular flow; whether a distinguished massless vector can be attached to the family is open. Third, the shape of the curve---the origin and location of its maximum and its precise large-mass asymptotics---invites an analytic treatment. What is settled is the structural point of this paper: an explicitly constructed, mass-indexed, wedge-localized family on which the mass-independent isolated-wedge modular generator produces a genuine, strictly positive, mass-dependent Araki relative entropy in $1{+}1$ and $1{+}d$ dimensions.

\section*{Acknowledgements}

\begin{sloppypar}
The authors acknowledge financial support from the Brazilian agencies CNPq, CAPES and FAPERJ. S.~P. Sorella, I.~Roditi and M.~S. Guimaraes are CNPq researchers, under the contracts 302991/2024-7 (S.P.S.), 319060/2025-0 (I.R.) and 309793/2023-8 (M.S.G.).
\end{sloppypar}

The authors also acknowledge the use of ChatGPT (OpenAI) and Claude (Anthropic) as auxiliary tools for discussion, organization and language refinement during the preparation of the manuscript. All scientific statements, derivations, checks and final wording were reviewed by the authors, who take full responsibility for the content.

\section*{Data Availability Statement}

No research datasets were generated or analyzed in this study. All numerical values and plots displayed in the manuscript were obtained directly from the analytical expressions derived in the text. The exact figure-generation script (\texttt{make\_figures.py}, Python/Matplotlib, scalable embedded fonts) is distributed with the source of this manuscript: it evaluates Eq.~\eqref{eq:entropydef} by spectral differentiation on $2^{15}$-point rapidity grids centered at the comoving point $\theta=-\log m$, cross-checks the result against the spectral representation \eqref{eq:spectralpositivity} (agreement to $15$ significant digits), and computes the transverse averages \eqref{eq:transverseaverage} by adaptive Gauss--Kronrod quadrature. For $m\gtrsim20$ the sinc oscillation frequency grows like $m^2$ and correspondingly finer grids are required.

\appendix

\section{Elementary derivation of the cutting projector}\label{app:cutting}

This appendix is deliberately elementary. It records the algebraic mechanism behind
\begin{equation}
P_H=(1+s_H)(1-\delta_H)^{-1} \label{eq:PHapp}
\end{equation}
on an explicitly stated spectral core on which all displayed inverse and Tomita--Takesaki operations are defined and bounded. Its purpose is to make the geometric action transparent, not to reproduce the full spectral-core and closure analysis of the unbounded cutting projection; for that rigorous operator-theoretic formulation---including the statement that the operator defined on the core below is closable with closure the cutting projection on its natural domain---we refer to the standard-subspace literature, in particular the cutting-projection treatment of Ciolli--Longo--Ruzzi \cite{CLR}. Within this deliberately restricted setting, the point is to see directly that the formula returns the same $H$-component appearing in the unique decomposition of a vector.

\subsection{The spectral core}\label{app:core}

Let $E(\cdot)$ be the spectral measure of $\delta_H$. For $\varepsilon\in(0,\tfrac12)$ define the inversion-symmetric set
\begin{equation}
\Lambda_\varepsilon=\bigl[\varepsilon,\,1-\varepsilon\bigr]\cup\bigl[(1-\varepsilon)^{-1},\,\varepsilon^{-1}\bigr],
\qquad Q_\varepsilon:=E(\Lambda_\varepsilon). \label{eq:cutoff}
\end{equation}
Since $j_H\delta_Hj_H=\delta_H^{-1}$ implies $j_HE(B)j_H=E(B^{-1})$ for Borel sets $B$, and $\Lambda_\varepsilon$ is invariant under $\lambda\mapsto\lambda^{-1}$, the projection $Q_\varepsilon$ commutes with $\delta_H$, with $j_H$, and hence with $s_H$ on $\dom s_H$. In particular
\begin{equation}
Q_\varepsilon H\subset H,\qquad Q_\varepsilon H'\subset H'. \label{eq:cutoffinv}
\end{equation}
On $\ran Q_\varepsilon$ the operators $\delta_H^{\pm1}$, $\delta_H^{\pm1/2}$ and $(1-\delta_H)^{-1}$ are all bounded, because the spectrum there is confined to $\Lambda_\varepsilon$, which is bounded away from $0$, $1$ and $\infty$.

Factoriality enters through the spectral point $1$. One has $H\cap H'=\{h\in H:\ \delta_Hh=h\}$ (combining $s_Hh=h$ with $s_{H'}h=h$ gives $\delta_Hh=h$; conversely $\delta_Hh=h$ and $h\in H$ give $j_Hh=h$, hence $h\in H'$), so $H$ is factorial precisely when $1$ is not an eigenvalue of $\delta_H$. Then $E(\{1\})=0$, $Q_\varepsilon\to1$ strongly as $\varepsilon\to0$, and the real-linear set
\begin{equation}
\mathcal D_0:=\bigcup_{\varepsilon>0}Q_\varepsilon\,(H+H') \label{eq:core}
\end{equation}
is dense in $H+H'$ and invariant under all the operators listed above. All computations in this appendix take place on $\mathcal D_0$; the statement that the operator so defined extends to the closed cutting projection on its natural domain is quoted from Ref.~\cite{CLR}, not reproved here.

\subsection{Geometric decomposition and factoriality}

Let $H$ be a factorial standard real subspace. On $\mathcal D_0$ write, for some $\varepsilon>0$,
\begin{equation}
\xi=h+j_Hk,\qquad h,k\in Q_\varepsilon H, \label{eq:appdecomp}
\end{equation}
since $j_HH=H'$ and $Q_\varepsilon$ commutes with $j_H$. Factoriality $H\cap H'=\{0\}$ makes this decomposition unique. The operator to be identified is the cutting projector itself,
\begin{equation}
P_H=(1+s_H)(1-\delta_H)^{-1}. \label{eq:appPH}
\end{equation}
Accordingly, it is enough to prove directly that
\begin{equation}
P_Hh=h\quad(h\in Q_\varepsilon H),\qquad P_Hj_Hk=0\quad(k\in Q_\varepsilon H). \label{eq:appgoals}
\end{equation}
These two properties reproduce exactly the geometric definition $P_H(h+j_Hk)=h$.

\subsection{Minimal modular identities}

We use
\begin{equation}
s_H=j_H\delta_H^{1/2},\qquad s_H^2=\mathbf 1,\qquad j_H\delta_Hj_H=\delta_H^{-1}, \label{eq:appidentities}
\end{equation}
which imply
\begin{equation}
s_H\delta_H=\delta_H^{-1}s_H,\qquad s_Hj_H=\delta_H^{-1/2}. \label{eq:appimplied}
\end{equation}
For $u\in H$, $s_Hu=u$, hence
\begin{equation}
j_Hu=\delta_H^{1/2}u. \label{eq:appjHu}
\end{equation}
On $\ran Q_\varepsilon$ the operator $1-\delta_H$ is bounded with bounded inverse, so all manipulations below are legitimate there; the possible spectral singularity at $1$ has been removed by the cutoff and reappears only in the closure statement quoted from Ref.~\cite{CLR}.

\subsection{\texorpdfstring{$P_H$}{PH} is the identity on \texorpdfstring{$H$}{H}}

Take $h\in Q_\varepsilon H$ and set
\begin{equation}
x=(1-\delta_H)^{-1}h,\qquad x-\delta_Hx=h. \label{eq:appx}
\end{equation}
Apply $s_H$:
\begin{equation}
s_Hx-\delta_H^{-1}s_Hx=h. \label{eq:appsx}
\end{equation}
Multiplying by $\delta_H$ and then by $-1$ gives
\begin{equation}
s_Hx-\delta_Hs_Hx=-\delta_Hh. \label{eq:appsx2}
\end{equation}
Adding this to $x-\delta_Hx=h$ yields
\begin{equation}
(1-\delta_H)(x+s_Hx)=(1-\delta_H)h. \label{eq:appadd}
\end{equation}
Since $1-\delta_H$ is injective on $\ran Q_\varepsilon$,
\begin{equation}
x+s_Hx=h. \label{eq:appconclusion1}
\end{equation}
Therefore
\begin{equation}
(1+s_H)(1-\delta_H)^{-1}h=h. \label{eq:appresult1}
\end{equation}

\subsection{\texorpdfstring{$P_H$}{PH} annihilates the complementary component}

Take $k\in Q_\varepsilon H$ and set
\begin{equation}
z=(1-\delta_H)^{-1}j_Hk,\qquad z-\delta_Hz=j_Hk. \label{eq:appz}
\end{equation}
Apply $s_H$:
\begin{equation}
s_Hz-\delta_H^{-1}s_Hz=\delta_H^{-1/2}k. \label{eq:appsz}
\end{equation}
Multiplying by $\delta_H$ gives
\begin{equation}
\delta_Hs_Hz-s_Hz=\delta_H^{1/2}k=j_Hk, \label{eq:appsz2}
\end{equation}
where the last equality uses $k\in H$. Thus
\begin{equation}
(1-\delta_H)s_Hz=-j_Hk. \label{eq:appsz3}
\end{equation}
Comparing with $(1-\delta_H)z=j_Hk$ and adding,
\begin{equation}
(1-\delta_H)(z+s_Hz)=0. \label{eq:appzero}
\end{equation}
Injectivity of $1-\delta_H$ on $\ran Q_\varepsilon$ gives
\begin{equation}
z+s_Hz=0, \label{eq:appconclusion2}
\end{equation}
and hence
\begin{equation}
(1+s_H)(1-\delta_H)^{-1}j_Hk=0. \label{eq:appresult2}
\end{equation}

\subsection{Conclusion and operator-algebraic meaning}

For $\xi=h+j_Hk\in\mathcal D_0$, the two properties established above give directly
\begin{equation}
P_H\xi=P_Hh+P_Hj_Hk=h. \label{eq:appfinal}
\end{equation}
Thus, on the spectral core of Eq.~\eqref{eq:core}, the modular expression has the geometric action of the cutting projector; its closure is the cutting projection on the natural domain \cite{CLR}. The derivation displays the geometric content of the formula: $(1+s_H)(1-\delta_H)^{-1}$ keeps precisely the $H$ component and removes the $H'$ component.

The factorial assumption is not decorative. The condition $H\cap H'=\{0\}$ is the one-particle counterpart of trivial center for the associated CCR von Neumann algebra, it is what makes the real decomposition unique, and, as shown in Appendix~\ref{app:core}, it is exactly the absence of the eigenvalue $1$ of $\delta_H$ that makes the spectral core dense. Thus the cutting projector reflects directly the factor structure of the local algebra. Domain questions remain essential because $P_H$ is generally unbounded and the spectral point $1$ of $\delta_H$ may obstruct a naive everywhere-defined inverse.

\section{An elementary second-quantized derivation of the one-particle formula for Araki relative entropy}\label{app:fock}

This appendix is a pedagogical calculation. It explains the elementary Fock-space mechanism behind the one-particle quadratic form used in Sec.~\ref{sec:oneparticle}. It does \emph{not} by itself derive Araki relative entropy from its original relative-modular-operator definition. The rigorous reduction of coherent-state relative entropy on CCR algebras to one-particle modular data is the content of the results in Refs.~\cite{CLR,CGP,BCD}.

Let $\Hh$ be the one-particle Hilbert space and let
\begin{equation}
\Gamma_s(\Hh)=\mathbb C\Omega\oplus\Hh\oplus\Hh^{\otimes_s2}\oplus\Hh^{\otimes_s3}\oplus\cdots \label{eq:fockspace}
\end{equation}
be the bosonic Fock space. For $h\in\Hh$, the normalized coherent vector is
\begin{equation}
W(h)\Omega=e^{-\|h\|^2/2}\sum_{n=0}^\infty\frac{h^{\otimes n}}{\sqrt{n!}}. \label{eq:coherentvector}
\end{equation}
Thus the coherent excitation has components in every particle-number sector; its $n$-particle component is $e^{-\|h\|^2/2}h^{\otimes n}/\sqrt{n!}$.

For a one-particle operator $A$, its additive second quantization acts on the $n$-particle sector as
\begin{equation}
d\Gamma(A)\big|_{\Hh^{\otimes_sn}}=\sum_{j=1}^n\mathbf 1\otimes\cdots\otimes A\otimes\cdots\otimes\mathbf 1,\qquad d\Gamma(A)\Omega=0. \label{eq:dGamma}
\end{equation}
On a symmetric tensor $h^{\otimes n}$, each of the $n$ summands gives the same scalar product. Hence
\begin{equation}
\langle h^{\otimes n},d\Gamma(A)h^{\otimes n}\rangle=n\,\langle h,Ah\rangle\,\|h\|^{2(n-1)}. \label{eq:sector}
\end{equation}
Orthogonality of distinct particle-number sectors now gives
\begin{align}
\langle W(h)\Omega,\,d\Gamma(A)\,W(h)\Omega\rangle
&=e^{-\|h\|^2}\sum_{n=1}^\infty\frac{n}{n!}\,\langle h,Ah\rangle\,\|h\|^{2(n-1)} \label{eq:focksum}\\
&=e^{-\|h\|^2}\,\langle h,Ah\rangle\sum_{r=0}^\infty\frac{\|h\|^{2r}}{r!}
=\langle h,Ah\rangle. \label{eq:fockresult}
\end{align}
Therefore, whenever the indicated expectation and quadratic form are defined,
\begin{equation}
\boxed{\ \langle W(h)\Omega,\,d\Gamma(A)\,W(h)\Omega\rangle=\langle h,Ah\rangle.\ } \label{eq:fockidentity}
\end{equation}

For the modular application, take $A=k_H=-\log\delta_H$ on its quadratic-form domain. Equation~\eqref{eq:fockidentity} then makes transparent the elementary second-quantization step behind
\begin{equation}
S_H(h)=-\langle h,\log\delta_H\,h\rangle=\langle h,k_Hh\rangle,\qquad h\in H. \label{eq:fockentropy}
\end{equation}
The nontrivial theorem is the identification of the relevant relative-modular expression for the coherent state with the second-quantized modular dynamics; the present appendix only isolates the subsequent Fock-to-one-particle calculation. This distinction is precisely why the cited CCR and standard-subspace results remain essential.

\section{Operational meaning of relative entropy}\label{app:operational}

Relative entropy quantifies distinguishability, but its asymptotic operational meaning must be stated with the correct testing problem. In asymmetric quantum hypothesis testing, where one error probability is constrained, the quantum Stein lemma \cite{OgawaNagaoka} identifies the relative entropy as the optimal exponential rate of the other error probability. By contrast, symmetric Bayesian minimum-error discrimination is governed asymptotically by the quantum Chernoff exponent \cite{Chernoff}, not directly by $S(\varphi\,|\,\omega)$. We use ``distinguishability'' in this precise modular/information-theoretic sense and do not identify the symmetric minimum error with $\exp[-NS]$.

\end{document}